\documentclass{iau}
\usepackage{graphicx}

\title[Kepler RR Lyrae stars: beyond period doubling] 
{Kepler RR Lyrae stars: beyond \\ period doubling}

\author[L. Moln\'ar, J.M. Benk\H{o}, R. Szab\'o and Z. Koll\'ath]   
{L. Moln\'ar$^{1,2}$, J.M. Benk\H{o}$^1$, R. Szab\'o$^1$, Z. Koll\'ath$^{2,1}$}

\affiliation{$^1$Konkoly Observatory, MTA CSFK \\ H-1121 Konkoly Thege Mikl\'os \'ut 15-17, Budapest, Hungary \\ email: {\tt molnar.laszlo@csfk.mta.hu} \\[\affilskip]
$^2$  Institute of Mathematics and Physics, Savaria Campus, University of West Hungary \\ H-9700 Szombathely, K\'arolyi G\'asp\'ar t\'er 4, Hungary}

\pubyear{2014}
\volume{301}  
\pagerange{1--2}
\jname{Precision Asteroseismology}
\editors{J.A. Guzik, W.J. Chaplin, G. Handler \& A. Pigulski, eds.}
\begin{document}

\maketitle

\begin{abstract}
We examined the complete short cadence sample of fundamental-mode \textit{Kepler} RR Lyrae stars to further investigate the recently discovered dynamical effects such as period doubling and additional modes. Here we present the findings on four stars. V450 Lyr may be a non-classical double-mode RR Lyrae star pulsating in the fundamental mode and the second overtone. For the three remaining stars we observe the interaction of three different modes. Since the period ratios are close to resonant values, we observe quasi-repetitive patterns in the pulsation cycles in the stars. These findings support the mode-resonance explanations of the Blazhko effect.
\keywords{stars: variable: RR Lyrae, Kepler}
\end{abstract}

\firstsection 
\section{Introduction}
The discoveries of space-based photometric missions such as \textit{Kepler}, \textit{CoRoT} and \textit{MOST} have reshaped our understanding of the pulsation of RR Lyrae stars. Not long ago, these stars were thought to be fairly simple, with three distinct classes: fundamental-mode, first-overtone and double-mode pulsators, with the only, though serious complicating factor being the Blazhko effect. It turned out, however, that the stars, especially the modulated ones show several dynamic phenomena: period doubling, additional modes and near-resonant states (Szab\'o, in this volume, and references therein). 

\section{Multi-mode RR Lyrae stars and their consequences}
Due to their sharpness, light maxima of RR Lyrae stars are not well covered by the 29.4 min long-cadence sampling of \textit{Kepler}. Therefore we analyzed only those RR Lyrae stars which have available short-cadence (SC) data. We searched for regularities and beating patterns in the light curves of all modulated RRab stars and compared them to the frequency content of the data. Beating can be separated from other proposed cycle-to-cycle irregularities (\cite[Gillet~2013]{gillet13}) and temporal evolution can be followed with the visual method. We found four different types of behaviour occurring as an interaction between four different modes: the fundamental (FM), the 1st, 2nd and 9th overtones (O1, O2 and PD). The modes labeled O1 and O2 are not necessarily radial modes in all cases but have similar frequency ratios in all stars. The 9th overtone has been observed only through period doubling when it is in resonance with the fundamental mode.

\begin{figure}[h]
\begin{center}
 \includegraphics[width=5.2in]{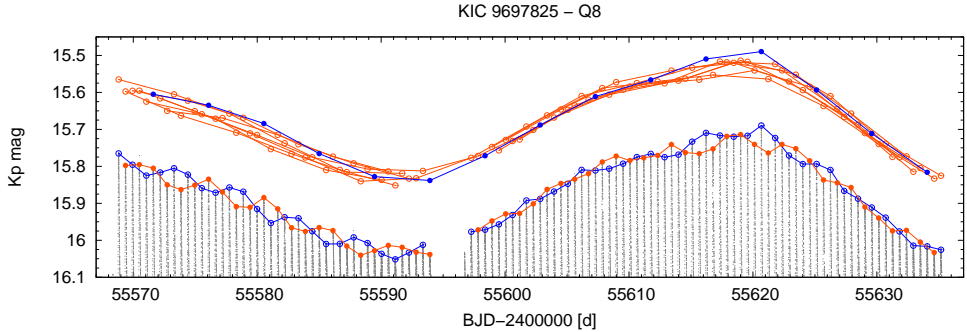} 
 \caption{Dots: the bright part of the light curve of V360 Lyr (KIC 9697825). Every second maxima are connected with filled points and open circles, respectively. If we connect every eighth maxima instead (shifted vertically), the curves become much smoother, revealing the true structure of the beating pattern. One realization is highlighted with filled points. }
   \label{fig1}
\end{center}
\end{figure}

\begin{itemize}
 \item FM + PD + O1 (RR~Lyr): the beating pattern of RR~Lyr is dominated by period doubling but beating with the first overtone sometimes creates quasi-repetitive patterns lasting six pulsation cycles with differing maximum brightnesses (\cite[Moln\'ar et al.~2012]{molnar12a}).
 \item FM + O1 + PD (V360~Lyr): the first overtone has much higher amplitude than period doubling. The FM-O1 frequency ratio is close to 8:11, resulting in the quasi-repetition of an 8 FM-period long beating pattern (Fig.~\ref{fig1}).
 \item FM + O2 (V460~Lyr): in this star only the FM and the 2nd overtone are visible, making it the third known FM/O2 double-mode star. The frequency ratio is close to 3:5, leading to a quasi-3-cycle beating pattern.
 \item FM + O2 + PD (KIC 7257008): the period ratio is the same as above but more complicated patterns are also observable. The frequency spectrum shows the signs of period doubling as well as peaks around the O1 mode but the latter are in fact $3f_0-f_2$ combination peaks.
\end{itemize}

Although other stars show more complicated patterns in the cycle-to-cycle amplitude variations that are harder to describe, almost all show similar additional frequencies. These findings support the mode-interaction model of the Blazhko effect (\cite[Nowakowski \& Dziembowski 2001]{nd01}; \cite[Buchler \& Koll\'ath 2011]{bk11}). On the other hand, similar cycle-to-cycle irregularities described in the shock-interaction model (\cite[Gillet 2013]{gillet13}) most likely arise simply from the combined effect of the beating of the modes and the uncertainties induced by the sparse sampling of the long cadence data instead of stochastic processes.

\section*{Acknowledgements}
This work has been supported by by the Hungarian OTKA grant K83790, the MB08C 81013 Mobility-grant of the MAG Zrt., the KTIA URKUT\_10-1-2011-0019 grant and the ``Lend\"ulet-2009" Young Researchers' Programme. The work of L. Moln\'ar was supported by the European Union and the State of Hungary, co-financed by the European Social Fund in the framework of T\'AMOP 4.2.4. A/2-11-1-2012-0001 `National Excellence Program'. R.Sz. was supported by the J\'anos Bolyai Research Scholarship. L.M. and R.Sz. acknowledge the IAU grants for the conference.

\end{document}